\documentclass[a4paper]{iopart}
\usepackage{graphicx}

\begin{document}

\title{Electronic states of laterally coupled quantum rings}

\author{J. Planelles$^1$, F. Rajadell$^1$, J.I. Climente$^{2,1}$, M. Royo$^1$ and J.L. Movilla$^1$}
\address{$^1$Departament de Ci\`encies Experimentals, UJI, Box 224, E-12080 Castell\'o, Spain}
\address{$^2$CNR-INFM S3, Via Campi 213/A, 41100 Modena, Italy}
\ead{josep.planelles@exp.uji.es}

\begin{abstract}

The conduction band electron states of laterally-coupled semiconductor quantum 
rings are studied within the frame of the effective mass envelope function theory.
We consider the effect of axial and in-plane magnetic fields for several inter-ring
distances, and find strong changes in the energy spectrum depending on the
coupling regime.
Our results indicate that the magnetic response accurately monitors
the quantum ring molecule dissociation process.
Moreover, the anisotropic response of the electron states to in-plane 
magnetic fields provides information on the orientation of the quantum 
ring molecule.

\end{abstract}



\section{Introduction}

Quantum rings (QRs) stand as an alternative to quantum dots (QDs)
as zero-dimensional structures for eventual use in nanotechnology devices.
The main differences between the physics of QRs and that of QDs follow 
from the doubly-connected geometry of the rings, which provides them 
with a characteristic electronic shell structure, magnetic field 
response and transport properties.\cite{ViefersPE,LeePE,NederPRL}
While much attention has been devoted in the last years to the study of 
'artificial molecules' made of coupled QDs (see e.g. Refs.\cite{Austing_book,
DiVicenzoSCI,JaskolskiIJQC} and references therein), 
only recently their QR counterparts have started being addressed.
A number of experimental and theoretical works have studied
vertically-coupled\cite{GranadosAPLSuarezNT,ClimentePRB,MaletPRB}, 
and concentrically-coupled\cite{ManoNLManoPRB,SzafranPRB,PlanellesEPJB,ClimentePRB2}
QRs.
Conversely, to our knowledge, laterally-coupled quantum rings 
(LCQRs) have not been investigated yet.
This is nonetheless an interesting problem: on the one hand, 
LCQRs constitute 'artificial molecules' with unique topology (two
LCQRs may be triply-connected), what should be 
reflected in unique energy structures; on the other hand, the 
formation of pairs of LCQRs in the synthesis of self-assembled QRs 
is apparent.\cite{LeeNTSchrammJCG} 
Therefore, one could investigate LCQRs experimentally by probing
spectroscopically the reponse of selected individual entities from a macroscopic
sample of self-assembled QRs, as done e.g. in Ref.\cite{WarburtonNAT}
for single QRs.

In this work we study the conduction band single-electron energy levels 
and wave functions of a pair of nanoscopic LCQRs, as a function of 
the distance between the two constituent QRs.
Particular emphasis is placed on the effect of external magnetic fields, 
 applied along the axial and two transversal in-plane directions, which
lead to characteristic magnetic responses depending on the strength
of the inter-ring coupling regime.

\section{Theoretical considerations}

Since usual QRs have much stronger vertical than lateral confinement,\cite{LeePE} 
we calculate the low-lying states of LCQRs using a two-dimensional
effective mass-envelope function approximation Hamiltonian which 
describes the in-plane ($x-y$) motion of the electron in the
ring. In atomic units, the Hamiltonian may be written as:

\begin{equation}
\label{eq1}
H=\frac{1}{2 m^*} (\mathbf{p}+\mathbf{A})^2+V(x,y)
\end{equation}

\noindent where $m^*$ stands for the electron effective mass
and $V(x,y)$ represents a finite scalar potential which confines
the electron within the lateral limits of the double ring heterostructure.
Here we define $x$ as the direction of dissociation of the LCQRs.
$\mathbf{A}$ is the vector potential, whose value depends on 
the orientation of the magnetic field $B$. 
Actually, the choice of $\mathbf{A}$ is limited by the
requirement that it should make it possible to separate $(x-y)$ coordinates
from $z$ in the Hamiltonian.\cite{PlanellesPE}
Within the Coulomb gauge, for a field applied along $z$ (axial magnetic field), 
this is fulfilled e.g. by $\mathbf{A_{B_z}}=(-y,x,0)\frac{1}{2} B$.
For an in-plane magnetic field applied along $x$ ($y$), this is fulfilled 
e.g. by $\mathbf{A_{B_x}}=(0,0,y) B$ ($\mathbf{A_{B_y}}=(0,0,-x) B$).
Replacing these values of the vector potential in Hamiltonian (\ref{eq1}) 
we obtain:

\begin{eqnarray}
\label{eq2}
H (B_z)=\frac{\hat p_{\parallel}^2}{2 m^*} + \frac{B_z^2}{8 \; m^*} (x^2+y^2)
- i \frac{B_z}{2 \; m^*} (x \frac{\partial}{\partial y}-y \frac{\partial}{\partial x}) + V(x,y),\\
\label{eq3}
H (B_x)=\frac{\hat p_{\parallel}^2}{2 m^*} + \frac{B_x^2 }{2 \; m^*} y^2 + V(x,y),\\
\label{eq4}
H (B_y)=\frac{\hat p_{\parallel}^2}{2 m^*} + \frac{B_y^2 }{2 \; m^*} x^2 + V(x,y).
\end{eqnarray}

The eigenvalue equations of Hamiltonians (\ref{eq2}-\ref{eq4}) are solved numerically using
a finite-difference method on a two-dimensional grid ($x,y$) extended far beyond the 
LCQR limits.  This discretization yields an eigenvalue problem of a huge asymmetric 
complex sparse matrix that is solved in turn by employing the iterative 
Arnoldi factorization.\cite{arnoldi}

In this work we investigate nanoscopic laterally-coupled GaAs QRs embedded 
in an Al$_{0.3}$Ga$_{0.7}$As matrix. We then use an effective
mass $m^*$=0.067 and a band-offset of $0.262$ eV.\cite{YamagiwaJL}
The pair of rings which constitute the artificial molecule have inner radius 
$r_{in}=12$ nm and outer radius $r_{out}=16$ nm, and the separation
between their centers is given by the variable $d$.

\section{Results and discussion}
\subsection{Zero magnetic field}

\begin{figure}[h]
\begin{center}
\includegraphics[height=13cm,angle=270]{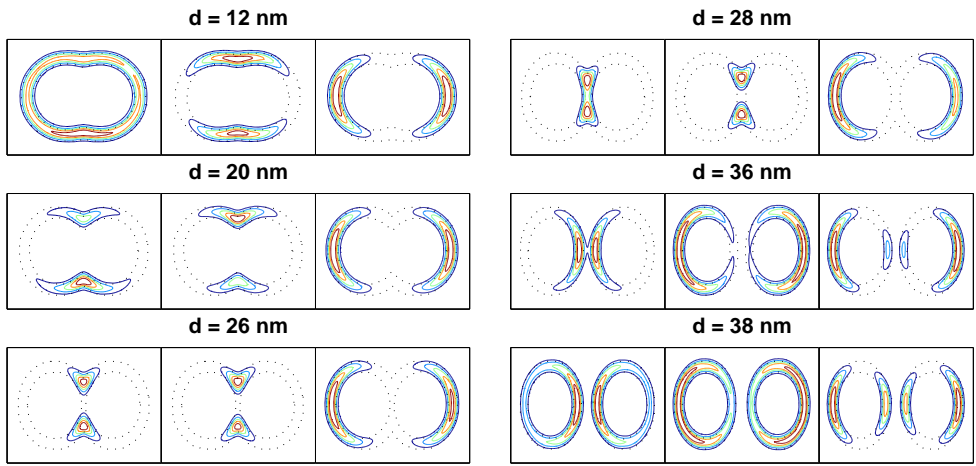}
\end{center}
\caption{(Color online). Contours of the wave functions corresponding to the three lowest-lying
electron states (from left to right) of LCQRs with different inter-ring 
distance $d$ at $B=0$. Dotted lines denote the confinement potential profile.}\label{Fig1}
\end{figure}

We start by studying the electron wave function localization in LCQRs
for increasing inter-ring distances, from the strongly coupled to the weakly 
coupled regime (a process we shall hereafter refer to as \emph{dissociation of 
the quantum ring molecule}), in the absence of external fields.
Figure \ref{Fig1} illustrates the wave functions of the three lowest-lying electron
states for several values of $d$. The corresponding profiles of the confining 
potential barrier are also represented using dotted lines. 
When the QRs are strongly coupled ($d=12$ nm), one clearly
identifies a $s$-like ground state and two $p$-like excited states.
The states localize along the arms of the LCQRs, as if in a single elliptical QR, 
with some excess charge deposited in the region where the two QRs overlap.
As the inter-ring distance increases, the available space in the overlapping
regions first increases. This tends to localize the ground and first 
excited states in such regions ($d=18$ nm) until they eventually become the 
even and odd solutions of a double quantum well ($d \sim 26$ nm). 
For further increased inter-ring distance, an inner arm of the LCQRs is formed. 
As a result, the ground state tends to localize along it ($d=28$ nm), thus 
benefiting from a reduced centrifugal energy.
Meanwhile, the first excited state, which is not so prone to minimize
centrifugal forces due to its $p$-like symmetry, prefers to spread along
the external arms of the rings.
Finally, for longer inter-ring distances, the QRs start detaching. When the
rings are close to each other, tunneling between the two structures is significant
and the ground state remains localized mostly in the middle of the LCQRs ($d=36$ nm), 
but it soon evolves into the ground state of single QRs, with a tunneling
acting as a small perturbation ($d=38$ nm). 
All along the dissociation process, the second and higher excited states
remain relatively insensitive to changes in $d$ due to their larger kinetic energy.

\subsection{Effect of external magnetic fields}

\begin{figure}[h]
\begin{center}
\includegraphics[height=13cm,angle=270]{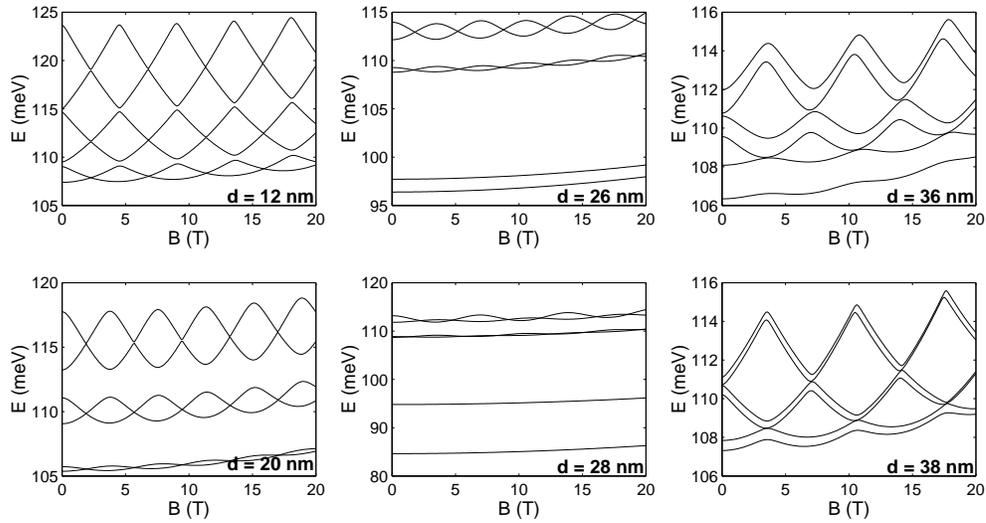}
\end{center}
\caption{Low-lying electron energy levels vs axial magnetic field in LCQRs with different inter-ring
distance.}\label{Fig2}
\end{figure}

We next study the response of the electron energy levels to external 
magnetic fields. In general, the magnetic response can be understood from
the $B=0$ charge distribution described in the previous section.
This is particularly clear in the case of an axial magnetic field, where
the field barely squeezes the wave functions shown in Fig.\ref{Fig1}.
To illustrate this, let us analyze Figure \ref{Fig2}, which depicts 
the low-lying energy levels against $B_z$ for several inter-ring 
distances. 
For $d=12$ nm, the picture resembles the usual Aharonov-Bohm spectrum of
single QRs,\cite{ViefersPE,PlanellesPE} save for the anticrossings appearing
between sets of two consecutive energy levels. These are due to the fact
that the electron states no longer have circular ($C_{\infty}$) symmetry
as in single QRs, but rather elliptical ($C_2$) one. 
Therefore, the pairs of eigenvalues which cross one another
correspond to the two irreducible representations 
of the $C_2$ symmetry group.  As $d$ increases and the confining potential
elongates, the anticrossing gaps become larger. Moreover, as the two
lowest-lying states tend to become the even and odd solutions of a double 
quantum well, they become nearly degenerate and,
being less efficient to trap magnetic flux, the amplitude of their 
energy oscillations is reduced (see panel corresponding to $d=20$ nm in Fig. 2).
In the next stage, around $d=28$ nm, the ground state localizes to
a large extent along the middle arm of the LCQRs and it takes essentially
a singly-connected shape, thus preserving a QD-like magnetic response.
On the other side, the first excited state tends to retrieve the $p$-like
symmetry. 
Therefore, its energy and magnetic behavior become similar to that of
the second QD excited state.
In the last stage, when the QRs are already detached and the ground
state wave function starts delocalizing among the two structures 
($d=36-38$ nm), the magnetic response is essentially that of a single 
QR with a perturbation arising from the tunneling between the rings, 
which rapidly diminishes with $d$. Notice that in this weak-coupling
limit, the period of the Aharonov-Bohm oscillations is larger than 
in the strongly coupled limit. This is due to the smaller area of the 
inner holes of the individual rings as compared to that of the 
strongly coupled structure.

\begin{figure}[h]
\begin{center}
\includegraphics[height=13cm,angle=270]{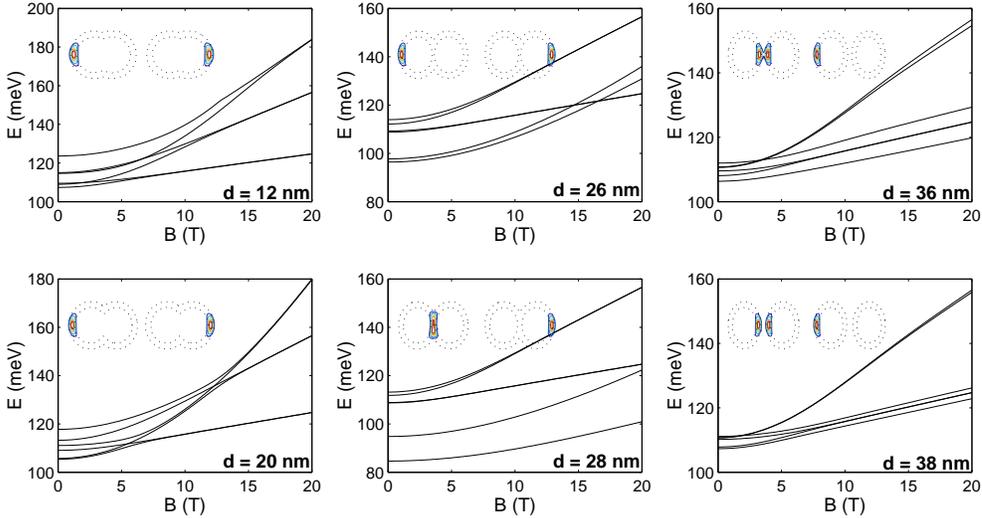}
\end{center}
\caption{(Color online). Low-lying electron energy levels vs in-plane magnetic field in LCQRs with different inter-ring
distance. The field is applied parallel to the dissociation axis. The insets show the wave 
functions of the two lowest-lying electron states (from left to right) at $B=20$ T.}\label{Fig3}
\end{figure}
 
\begin{figure}[h]
\begin{center}
\includegraphics[height=13cm,angle=270]{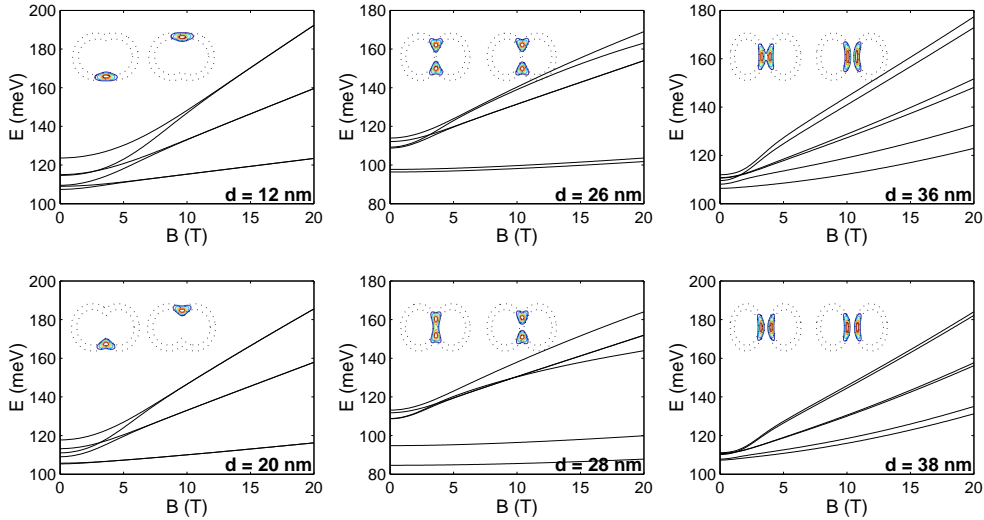}
\end{center}
\caption{(Color online). Low-lying electron energy levels vs in-plane magnetic field in LCQRs with different inter-ring
distance. The field is applied perpendicular to the dissociation axis. The insets show the wave 
functions of the two lowest-lying electron states (from left to right) at $B=20$ T.}\label{Fig4}
\end{figure}

Figures \ref{Fig3} and \ref{Fig4} show the energy levels against
in-plane magnetic fields directed along the $x$ and $y$ directions,
respectively. $B_x$ is applied along the dissociation axis and
it tends to squeeze the electron wave function in the $y$ direction.
The opposite holds for $B_y$. 
In both cases, when the QRs are strongly coupled ($d=12-20$ nm)
the effect of the field is to form pairs of degenerate energy levels. 
These are the even and odd solutions of double quantum wells building up 
in the longitudinal ($B_x$) or transversal ($B_y$) edges of the ring 
structure, as illustrated in the insets of the figures for the 
lowest-lying states. The formation of double well solutions 
at moderate values of in-plane magnetic fields has been also
reported for nanoscopic single QRs.\cite{PlanellesPE}
An additional feature is however present in LCQRs, because
the asymmetric confinement in the $x$ and $y$ directions leads
to anisotropic magnetic response. As a result, for instance,
one observes that the values of the field at which the double quantum 
well solutions are obtained are much smaller for $B_y$ than for $B_x$.
Thus, the two lowest-lying states at $d=12$ nm become degenerate at
$B \sim 5$ T for $B_y$, while they do so at $B \sim 8$ T for $B_x$.
Another difference in the spectrum is the presence of crossings
between given energy levels for $B_x$ (e.g. between the first and second 
excited states), which are missing for $B_y$. This is because the
two $p$-like states of the strongly coupled QRs are non-degenerate at
zero magnetic field, due to the eccentricity of the LCQR system, and 
the applied field may reverse their energy order depending on the 
direction.
When the coupling between the LCQRs is intermediate ($d \sim 26$ nm), 
the states which at $B=0$ have singly-connected wave functions 
behave as in QD, i.e. they depend weakly on the external field.
On the contrary, the doubly-connected states keep on behaving as in a QR,
i.e. they tend to form double quantum well solutions. 
Finally, for weakly coupled QRs ($d \sim 38$ nm), different limits 
are reached depending on the in-plane magnetic field direction. 
$B_y$ strongly enhances tunneling between the two QRs, so that 
the lowest-lying states are double well solutions mostly localized in the
vicinity of the tunneling region (see insets in Fig.~\ref{Fig4}). 
Conversely, for $B_x$ double well solutions localized either in the 
inner or in the outer edges of the QRs alternate (see insets in
Fig.~\ref{Fig3}).

\section{Conclusions}

We have studied the electron states of nanoscopic LCQRs as a 
function of the inter-ring distance and external magnetic fields.
The wave function localization at $B=0$ changes dramatically
depending on the inter-ring distance, and this gives rise to
characteristic magnetic responses for strong, intermediate and
weak coupling regimes. Moreover, a clearly anisotropic response
is found for in-plane fields applied parallel or perpendicular
to the LCQRs dissociation direction.
These results suggest that probing spectroscopically the 
magnetic response of electrons in LCQRs may provide valuable 
information on the strength of coupling and the orientation of 
the QR molecule. 


\ack

Financial support from MEC-DGI project CTQ2004-02315/BQU and UJI-Bancaixa 
project P1-B2002-01 is gratefully acknowledged.
J.I.C. has been supported by the EU under the Marie Curie IEF project 
MEIF-CT-2006-023797 and the TMR network ``Exciting''.


\end{document}